# ACCOUNTING FOR SPATIAL CORRELATION IN THE SCAN STATISTIC[1]

By Ji Meng Loh and Zhengyuan Zhu

*Columbia University and University of North Carolina at Chapel Hill*

The spatial scan statistic is widely used in epidemiology and medical studies as a tool to identify hotspots of diseases. The classical spatial scan statistic assumes the number of disease cases in different locations have independent Poisson distributions, while in practice the data may exhibit overdispersion and spatial correlation. In this work, we examine the behavior of the spatial scan statistic when overdispersion and spatial correlation are present, and propose a modified spatial scan statistic to account for that. Some theoretical results are provided to demonstrate that ignoring the overdispersion and spatial correlation leads to an increased rate of false positives, which is verified through a simulation study. Simulation studies also show that our modified procedure can substantially reduce the rate of false alarms. Two data examples involving brain cancer cases in New Mexico and chickenpox incidence data in France are used to illustrate the practical relevance of the modified procedure.

**1. Introduction.** Detection of clusters in spatial point processes is of great practical importance in a broad range of disciplines, such as epidemiology, astronomy and forestry, and has generated considerable interest among statisticians in recent years. See Lawson and Denison (2002) for a review of the diverse approaches to this problem. In recent literature spatial scan statistics have been widely used in studies of disease clustering in epidemiology and health science [e.g., Hjalmars et al. (1996), Viel et al. (2000), Sankoh et al. (2001) and Perez et al. (2002)]. The scan statistic was first studied by Naus (1965) and others as a means to detect clusters in a one-dimensional point process. The basic idea is to move a window with width $w$ along the interval over which the point process is observed. The

Received February 2007; revised July 2007.

[1]Supported in part by NSF Grants AST-05-07687 and DMS-06-05434 and SAMSI. Both authors contributed equally to the work and writing of this paper.

maximum number of points over all possible windows is recorded and used to test the null hypothesis of a purely random Poisson process. Kulldorff (1997) extended the scan statistic to the spatial setting, allowing for the detection of clusters of a multi-dimensional point process.

The spatial scan statistic is based on the likelihood ratio test: for a region $C$, a likelihood ratio test statistic is computed for testing the null hypothesis of equal rates within and outside $C$ versus the alternative of higher rates inside $C$. The spatial scan statistic is the likelihood ratio test statistic maximized over all possible $C$ (perhaps limited to, say, half the size of the observation region). A $p$ value for the cluster $C$ with the maximum value of the likelihood ratio test statistic is obtained by comparing the value of the spatial scan statistic for that dataset with the distribution under the independent Poisson or binomial model. Since the exact distribution of the test statistic cannot be easily determined analytically, it is approximated by the Monte Carlo simulation. This has been implemented in the SatScan software [Kulldorff et al. (1998b)].

In this work we examine the behavior of the spatial scan statistic when there is underlying overdispersion and spatial correlation in the data, and propose a modified spatial scan statistic to account for the overdispersion and spatial correlation. Overdispersion is commonly found in count data fitted with Poisson models [see Cox (1983), Breslow (1984), Lawless (1987) and McCullagh and Nelder (1989)]. For typical disease counts data in spatial epidemiology studies, it is also reasonable to expect some positive correlation between nearby locations. Both are evident in the real data examples we consider here.

We first look at this problem from a theoretical perspective, and show that when overdispersion or spatial correlation is present, the classical spatial scan statistic tends to produce more false positives than the nominal significance level asymptotically under certain conditions. This is confirmed by our simulation studies, which show that the classical spatial scan statistic identifies too many clusters when overdispersion or spatial correlation is present.

We provide a simple modification to the spatial scan statistic procedure to adjust the $p$ values of identified clusters (Section 2.3). Essentially, we fit a spatial generalized linear mixed model to the data which specifically includes a spatial component. This spatial component is designed to capture any spatial correlation or overdispersion in the data. In the Monte Carlo simulation to obtain the distribution of the test statistic, we use this new model rather than the independent Poisson model to simulate the distribution of the scan statistic. In our simulation study we find that, by using this simple modification, we substantially reduce the number of false positives. Details of our simulation study are given in Section 3.

For the purpose of illustration, we apply both the classical spatial scan statistic and our adjusted version to the New Mexico brain cancer data [Kulldorff et al. (1998b)]. A cluster which is identified by the classical spatial scan statistic as highly significant is not significant using our adjusted scan statistic. Our result is consistent with the result in Kulldorff et al. (1998b) which used extra covariates in the model. We also applied these methods to chickenpox incidence data in France. Here, we find that the classical spatial scan statistic identifies an excessive number of hot spots (one in almost every time period), while the adjusted scan statistic yields more reasonable results (see Section 4).

Section 2 introduces the notation, reviews the classical spatial scan statistic, presents the main theoretical results, and describes the algorithm for the adjusted scan statistic. Section 3 gives the simulation results and the data examples are in Section 4. The proof of Proposition 2 is relegated to the Appendix.

**2. Spatial scan statistic.** We begin by introducing some notation and the basic statistical model. Let $G = \bigcup_{i=1}^{m} A_i$ be the study region, where $\mathcal{A} = \{A_i,\ i = 1, 2, \ldots, m\}$ is a partition of $G$, that is, the study region $G$ is divided into $m$ nonoverlapping subregions $A_i$. At each subregion $A_i$, we observe $Y_{i,t}$, the number of cases of disease during time interval $t$, and the covariates $\mathbf{x}_{i,t} = (x_{i,t}^1, \ldots, x_{i,t}^p)'$. Examples of covariates include $N_{i,t}$, the baseline population at risk in $A_i$, age, race, gender distribution etc. Poisson and Bernoulli models are two typical models for $Y_{i,t}$. In this paper we focus on the Poisson models. Similar results can be derived for Bernoulli models as well. We are interested in detecting clusters of subregions for which the number of observed cases of disease is significantly higher than that predicted from the model.

2.1. *The classical spatial scan statistic.* Kulldorff (1997) introduced a spatial scan statistic for the detection of clusters where the null hypothesis is that the baseline process is an inhomogeneous Poisson process. We will refer to this null model as the following:

Model I: (inhomogeneous Poisson process/log-linear model)

$$Y_{i,t} | \lambda_{i,t} \sim \text{indep. Poisson}(\lambda_{i,t}),$$
(1)
$$\log \lambda_{i,t} = \log \int_{A_i} \lambda_{s,t}\, ds = \beta + \mathbf{x}_{i,t}\gamma + \log N_{i,t}.$$

For fixed $t$, we are interested in testing whether $\beta$ is a constant across all the subregions. A cluster is defined to be a region $C$, consisting of one or more of the $A_i \in \mathcal{A}$, within which $\beta$ is higher. For a given fixed $C$, testing whether $C$ is a cluster is just the generalized likelihood ratio test [e.g., Rice

[1995)] with the test statistic given by

$$
(2) \qquad LR_C = \left(\frac{Y_C}{N_C}\right)^{Y_C} \left(\frac{Y_G - Y_C}{N_G - N_C}\right)^{Y_G - Y_C},
$$

where $Y_G$ and $N_G$ are the total number of cases and population at risk, $Y_C = \sum_{A_i \subset C} Y_i$ and $N_C = \sum_{A_i \subset C} N_i$ are the number of cases and population at risk in region $C$. Kulldorff suggested computing $LR_C$ for all $C \subset \mathcal{A}$ to find $LR^* = \max_C LR_C$ over all possible subsets, and use it as the test statistic. The distribution of this test statistic is computed using Monte Carlo simulation under Model I, conditional on the total number of cases observed. Kulldorff, Tango and Park (2003) compared the spatial scan statistic with several other cluster detection methods, and concluded that the spatial scan statistic has an advantage for localized hotspot type clusters.

2.2. *Spatial scan statistic with spatial correlation*: *theoretical results.* The spatial scan statistic in its original setup (Model I) assumes that when there are no clusters, the numbers of disease cases in the individual subregions have Poisson distributions with rates that are spatially independent. Both the Poisson distribution and the spatial independence assumptions may be violated in some practical problems. In the context of disease surveillance, it is not unusual for the variability of the disease cases to be larger than those predicted by the Poisson distribution (over-dispersion). Spatial correlation is often present as well, due to the contagious nature of the disease, or to some latent variables that are related to the disease but were not included in the data collection or in the model. In what follows we present a null model that includes spatial correlation, which we refer to as Model II:

Model II: (Spatial GLMM/GSLM)

$$
(3) \qquad \begin{aligned} Y_{i,t}|\lambda_{i,t} &\sim \text{indep. Poisson}(\lambda_{i,t}), \\ \log \lambda_{i,t} &= \beta + \mathbf{x}_{i,t}\gamma + \log N_{i,t} + Z_{i,t}, \end{aligned}
$$

where $Z_{i,t}$ is a Gaussian process with covariance given by $\text{Cov}(Z_{i,t}, Z_{i',t'}) = C_\theta(s_i - s_{i'}, t - t')$, $s_i$ and $s_{i'}$ are the centers of subregions $i$ and $i'$, and $C_\theta$ is a positive definite function with parameter $\theta$, which may be a vector.

Model II is a special case of the model-based geostatistics model introduced in the seminal work of Diggle et al. (1998). It is also referred to in the literature as the spatial generalized linear mixed model, or generalized spatial linear model [Zhang (2002) and Christensen and Ribeiro Jr. (2002)]. Wikle (2002) used this model for a breeding bird survey dataset and addressed some of the computational issues on applying this model to large datasets. The role of the $Z_{i,t}$ term in (3) is to capture any (residual) spatial correlation in the data. The range of the correlation, smoothness of $Z$ as well as anisotropy, can be modeled with different choice of $C_\theta$.

As an aside, note the subtle differences between Model II and the following log-Gaussian Cox process model [Møller, Syversveen and Waagepetersen (1998)]:

Model III: (Cox process)

$$Y_{i,t}|\lambda_{i,t} \sim \text{indep. Poisson}(\lambda_{i,t}),$$

$$\lambda_{i,t} = \int_{A_i} \lambda_{s,t}\, ds, \log \lambda_{s,t} = \beta + \mathbf{x}_{i,t}\gamma + \log N_{i(s),t} + Z_{s,t},$$

where $Z_{s,t}$ is a Gaussian process with covariance given by $C_\theta(s-s', t-t')$. Here we have $\log \lambda_{i,t} = \beta + \mathbf{x}_{i,t}\gamma + \log N_{i,t} + Z_{i,t}$, where $Z_{i,t} = \log \int_{A_i} \exp\{Z_{s,t}\}\, ds$ does not have a Gaussian distribution. Model III may be conceptually more appealing based on the principle of invariance under scaling [Gelman (1996)], but Model II is easier to fit for the type of aggregated disease surveillance data we intend to model, and will be the model we use for the rest of the paper. We note that in this paper we assume that the $A_i$'s are pre-specified, as they usually are by the data collection process. In the case when the exact location of each case is known, it will be more desirable to model the point process of cases directly, in which case Model III will be a more appropriate model. The method we propose to modify the spatial scan statistic can be applied using Model III, if so desired. The only differences in the method would be in the way the parameters are estimated and in the simulation of the reference distribution.

In the rest of this section we restrict our attention to pure spatial processes and drop the subscript $t$ for brevity. All the results are applicable with the added time dimension. For simplicity of notation, we ignore covariates in the theoretical derivation below. The generalization to include covariates $\mathbf{x}_{i,t}$ is trivial.

Let $A = \bigcup_{i \in I} A_i$ be a subset of $\mathcal{A}$, $I \subset \{1, 2, \ldots, m\}$, $Y_A = \sum_{i \in I} Y_i$ be the observed counts in $A$, with $Y_i$ satisfying (3), $\lambda_A = \sum_{i \in I} \lambda_i$, and $\bar{\lambda}_A = \mathrm{E}[\lambda_A]$. We define $P_k^{(1)}$ and $P_k^{(2)}$ by

$$P_k^{(1)} = \Pr(Y_A = k | \bar{\lambda}_A),$$

$$P_k^{(2)} = \Pr(Y_A = k) = \mathrm{E}[\Pr(Y_A = k | \lambda_A)],$$

where the expectation is taken over $\lambda_A$.

PROPOSITION 1. *There exists a constant $K$ such that, for any $k > K$, $\sum_{j=k}^{\infty} P_j^{(2)} > \sum_{j=k}^{\infty} P_j^{(1)}$.*

PROOF. Let $\mu_1(dx)$ be the measure that has mass $(x!)^{-1}$ at $x = 0, 1, \ldots$, and $\mu_2(d\lambda_A)$ be the probability measure of $\lambda_A$. Furthermore, let

$$f_1(x) = \exp\{x \log \bar{\lambda}_A - \bar{\lambda}_A\},$$

$$f_2(x) = \int \exp\{x \log \lambda_A - \lambda_A\} \mu_2(d\lambda_A).$$

It is easy to check that

$$\int x f_1(x) \mu_1(dx) = \int x f_2(x) \mu_1(dx) = \bar{\lambda}_A.$$

Theorem 1 in Shaked (1980) implies that $f_2 - f_1$ has two sign changes, and the sign sequence is $+, -, +$. Proposition 1 follows as

$$\sum_{j=k}^{\infty} (P_j^{(2)} - P_j^{(1)}) = \int_k^{\infty} [f_2(x) - f_1(x)] \mu_1(dx). \qquad \square$$

REMARK. Proposition 1 shows that a mixture of Poisson distributions has a heavier right tail than the Poisson distribution with the same mean. It was used as an example in Shaked (1980). In the context of spatial scan statistics, it implies that if Model II is the truth, and the significance level is small enough, Model I gives a $p$ value which is on average smaller, resulting in more false positives.

PROPOSITION 2. *For* $\mathrm{Var}(Z_i) = \frac{\sigma^2}{n}$, *as* $n \to \infty$,

$$(4) \qquad \sum_{j=k}^{\infty} P_j^{(2)} = \sum_{j=k}^{\infty} P_j^{(1)} + (P_{k-2}^{(1)} - P_{k-1}^{(1)}) e^{2\beta} \frac{V_n}{2n} + O_p(n^{-3/2}),$$

*where* $V_n = n \, \mathrm{Var}(\sum N_i Z_i) = O(1)$.

The proof of Proposition 2 is provided in the Appendix. The following corollary follows directly from Proposition 2.

COROLLARY 1. $\sum_{j=k}^{\infty} P_j^{(2)}$ *is asymptotically larger than* $\sum_{j=k}^{\infty} P_j^{(1)}$ *iff* $k > \lambda + 1$.

REMARK. Corollary 1 gives a specific condition on the critical value ($k > \lambda + 1$) under which Model I gives a smaller $p$ value even when the variance of the random effects $Z_i$ is very small. It implies that the classical spatial scan statistic test procedure would be more likely to reject $H_0$ at any reasonable significance level if Model II is correct, resulting in more false positives. From Proposition 2, it is also clear that positive correlation leads to more extreme $p$ values for Model I and more false positives.

2.3. *Spatial scan statistic with spatial correlation*: *algorithm.*   If Model II is the appropriate null hypothesis, the classical spatial scan statistic (which assumes Model I) tends to give smaller $p$ values, resulting in more false alarms. A remedy for this is to obtain the distribution of the spatial scan statistic by simulation under Model II. We propose the following algorithm to appropriately account for the spatial correlation in the computation of $p$ values for the spatial scan statistic. It is only moderately more time consuming than the classical spatial scan statistic algorithm.

1. Use the regular spatial scan statistic to identify regions in space and time where there are no clusters.
2. Estimate parameters in Model II using data that excludes clusters.
3. Simulate M realizations of the process under Model II using the estimated parameters from step 2, and compute $LR^*$ for each realization to form a distribution of $LR^*$ under Model II.
4. Compute $LR^*$ for the data and find clusters in the data using the distribution of $LR^*$ from step 3.
5. Repeat steps 2–4 until the clusters in steps 2 and 4 are similar.

Parameter estimation for Model II is nontrivial, as the likelihood function is not in closed form: it involves a high-dimensional integral over random effects $Z_i$. We use the Bayes inference method for the generalized spatial linear model implemented in the R package geoRglm [Christensen and Ribeiro Jr. (2002)]. The Bayes method assumes a prior distribution on the parameters $(\beta, \theta)$, and give samples from the posterior distribution of $(\beta, \theta)$ conditional on the data using MCMC. More details on the choice of prior and correlation models are given in the simulation studies and data examples.

To simulate data $Y_i$ from Model II, one needs to first simulate $Z_i$, a Gaussian process with covariance given by $C_\theta$. Let $\Sigma$ be the covariance matrix of $\mathbf{Z} = (Z_1, \ldots, Z_m)'$, and $\Sigma = LL'$ be its Cholesky decomposition. $Z_i$ can be simulated by multiplying $L$ and a vector of independent normal random variables. To simplify the computation, we simulate $Z_i$ using the covariance matrix computed from $C_{\hat{\theta}}$, where $\hat{\theta}$ is the mean of the posterior distribution of $\theta$.

We would like to note here similar ideas presented in Efron (2004, 2007). In these two papers Efron considered simultaneous testing with a large number of tests, in the context of using false discovery rate (FDR) in genetic studies. Efron (2004) shows that the choice of different null hypotheses can yield different test decisions, and argues for the use of what he calls the empirical null to approximate the correct null. In Efron (2007) he examines the effect of correlation of test statistics on simultaneous testing procedures, and shows that the presence of correlation can affect the null and have substantial impact on the results of simultaneous testing. He recommends accounting for this effect, for example, with the use of the empirical null. An important assumption made in these two papers is that the number of interesting cases is small: about 1–5%, but not more than 10%.

The method presented in this work is in the same spirit as Efron (2004, 2007). Searching for clusters involves multiple testing of overlapping regions, so that these tests are correlated. The regular spatial scan statistic procedure accounts for the correlation in the overlapping regions by using the Monte Carlo simulation to approximate the distribution of the scan statistic. In this work we show that when there is spatial correlation in the data itself, the independent Poisson model used in the original spatial scan statistic does not produce the appropriate null hypothesis. By modeling the spatial correlation, and incorporating this in the Monte Carlo simulation, a more appropriate null distribution is used in the testing, and reduces the number of false positives. One difference from Efron (2004, 2007) is that instead of using the data to obtain an empirical null, we are proposing a parametric model to model the spatial correlation.

REMARK 1. The above algorithm essentially uses a parametric bootstrap approach. We do not have any theoretical results showing consistency of the approach. However, we found empirically that this procedure does help address the issue of excessive false alarms with the original spatial scan statistics (see Sections 3 and 4).

REMARK 2. As in Efron (2004, 2007), we require that the number of interesting clusters be small, say, at most 10% of detected clusters. This is usually the case in applications of disease surveillance. If many clusters are expected to be found, sophisticated statistical methods are probably not needed. Accordingly, we suggest that the significance level for step 1 of the algorithm be set to 0.1.

REMARK 3. When there is spatial correlation, instead of using the scan statistic, $LR^* = \max_C LR_C$, with $LR_C$ defined in (2), ideally $LR_C$ should be re-defined to include the modeled spatial correlation. However, this is often computationally prohibitive, since to evaluate the likelihood one would need to compute multiple integrals involving the nuisance parameters $Z_i$. Our proposed method is a simple, if ad hoc, procedure to improve the performance of the spatial scan statistic in the presence of spatial correlation.

**3. Simulation study.** Our simulation study consists of two parts. First, we investigate how the classical spatial scan statistic behaves when there is underlying correlated error in the true model. Second, we study the effectiveness of the modified procedure in reducing the number of false positives found by the spatial scan statistic.

We base our simulations around the New Mexico brain cancer dataset considered in Kulldorff et al. (1998a). In particular, we use the counties in this dataset as the regions for our simulation. We also use the 1991 population numbers. To keep things simple, we do not include any covariates in this simulation. Specifically, we model the number of disease cases $\mathbf{Y} = (Y_1, \ldots, Y_m)$ using Model II, with $\mathbf{Z} = (Z_1, \ldots, Z_m) \sim N(\mathbf{0}, \Sigma)$. The quantity $\beta$ in (3) is related to the incidence rate of the disease.

We take $Z$ to be a realization of a Gaussian random field with the Matérn covariance function, that is,

$$(5) \quad \text{Cov}(Z_i, Z_j) = \frac{\sigma^2}{2^{\nu-1}\Gamma(\nu)} \left(\frac{\nu^{1/2} d_{ij}}{\rho}\right)^\nu \mathcal{K}_\nu\left(\frac{d_{ij}}{\rho}\right),$$

where $d_{ij}$ is the distance between the centers of regions $i$ and $j$, $\Gamma$ is the gamma function and $\mathcal{K}_\nu$ is a modified Bessel function [Abramowitz and Stegun (1965)]. The parameter $\nu$ is related to the smoothness or differentiability of the random field with larger $\nu$ corresponding to a smoother field, while the parameter $\rho$ is related to the range of dependence in the values of the $Z_i$'s, with larger $\rho$ indicating stronger dependence. Finally, $\text{Var}(Z_i) = \sigma^2$.

For a set of values of $\nu$, $\rho$ and $\sigma$, we first generate a realization of $\mathbf{Z}$ at the locations of the counties. Each $Y_i$ is then simulated from a Poisson distribution with mean $\lambda_i$, with $\lambda_i$ given by (3). Next, we use SatScan to compute the scan statistic of the simulated values of $\mathbf{Y}$ to identify clusters and the $p$ value of the main cluster is recorded. A small $p$ value would indicate that the cluster identified by the scan statistic is significant. This whole procedure is repeated 1,000 times, each time with new values of $\mathbf{Y}$ and $\mathbf{Z}$, yielding 1,000 $p$ values corresponding to the main cluster found in each of the 1,000 simulated datasets. We then compute the proportion of these $p$ values that are smaller than a given significance level $\alpha$. If the datasets were simulated from Model I, the model assumed for the classical spatial scan statistic (Poisson, without the error $\mathbf{Z}$), then on average $\alpha \times 100\%$ of the $p$ values will be less then $\alpha$, that is, the type I error (false alarm rate) is the same as the significance level.

We ran the entire simulation described above using a range of values of $\nu, \rho$ and $\sigma$ and considered significance level $\alpha$ equal to 0.01, 0.05 and 0.1 to study the relationship between the false alarm rate and these parameters. The results are plotted in Figure 1. Our results are not affected by changes in the values of $\nu$, so we only show results for $\nu = 1$.

Figure 1 contains three plots. From left to right, these plots show the proportion of $p$ values that are less than 0.01, 0.05 and 0.1, respectively. In each plot, the 10 lines from top to bottom correspond to $\sigma = 0.1, 0.09, 0.08, \ldots, 0.01$, respectively. The dots at $\rho = 0$ give the proportions under the model assumed in the classical scan statistic, that is, with $\mathbf{Z} \equiv 0$.

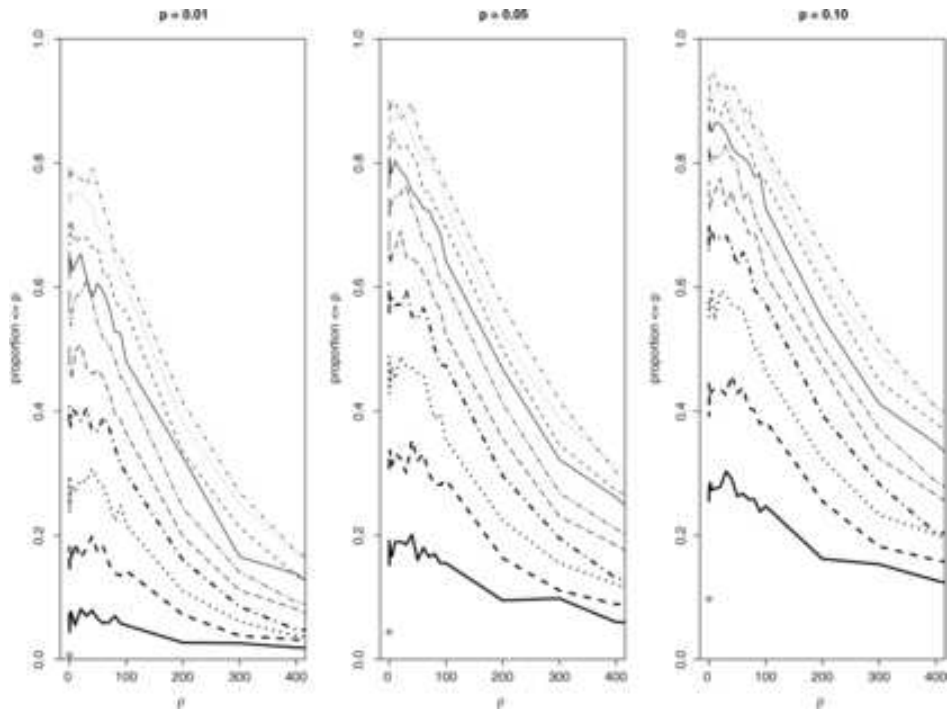

Fig. 1. *Plots of the proportion of p values that are 0.01, 0.05 and 0.1 (left to right) or less, obtained by applying the classical spatial scan statistic to data simulated under Model* II. *Parameters for Model* II *used are the following: $\rho$ from 0 to 400 (x-axis) and $\sigma = 0.01, 0.02, \ldots, 0.1$ (bottom line to top line).*

The behavior shown in Figure 1 can be understood by examining the properties of **Z** assumed in each simulation run. When $Z \equiv 0$, we find that the scan statistic behaves as it should: the proportion of $p$ values less than or equal to 0.01, say, is about 0.01.

When $Z \neq 0$, by Propositions 1 and 2, the scan statistic tends to find more clusters in the data. Figure 1 shows that this is indeed the case: the proportion of $p$ values less than or equal to 0.01, say, can be as high as 80%, indicating a severe problem with false alarms. This of course depends on the variance of **Z**. When the variance is large, the values of **Z** vary a lot more, causing the scan statistic to identify a lot more clusters in the data. As the variance of $Z$ is reduced, the proportion of false alarms is also reduced. However, we find that even for $\sigma = 0.01$, the proportion of $p$ values less than, say, 0.01 (left plot of Figure 1) can be as high as 0.07.

When the dependence between the values of $Z_i$ is very strong, that is, when $\rho$ is very large, the values of all $Z_i$ become very similar due to the strong dependence. This may produce an elevated level of risk of disease uniformly over all the counties, but with less variability, thus reducing the number of false alarms found by the spatial scan statistic.

There is a peak in the proportion at some small value of $\rho$, different from zero. For even smaller values of $\rho$, the proportion drops slightly, but even in the case of independent $Z_i$, we find that the proportion of $p$ values that are less than $p$ is still larger than the value that we would expect if the model assumed in the scan statistic is true.

Thus, the scan statistic is sensitive to model misspecification, specifically of the type where there is some error in the model involving $\lambda$, producing overdispersion or spatial correlation in $\mathbf{Y}$ relative to the Poisson model [see, e.g., McCullagh and Nelder (1989)]. Our simulation studies indicate that the behavior of the scan statistic $\mathbf{Z}$ depends a lot on the variability, as well as the range of correlation of $\mathbf{Z}$. We expect this type of error to occur frequently in real settings, through spatial correlation in the data and various sources of measurement errors. Some spatial correlation may be captured by incorporating covariates in the model. However, there may still be residual spatial correlation, and measurement errors in the covariates. Furthermore, there may be instances where covariates are not recorded in the data.

By modeling the overdispersion and spatial correlation present in a dataset, the modified algorithm given in Section 2.3 aims to reduce the number of false positives found by the classical spatial scan statistic. In the context of our simulation study, the effect will be to reduce the proportion of $p$ values that are less than the nominal value. The simulation experiments described below demonstrate the effectiveness of our algorithm.

For each of the 100 previously simulated realizations corresponding to each parameter set, we use Model II to model the spatial correlation. We then use the estimated parameters to generate 999 new realizations. The spatial scan statistic is computed for each of these new realizations. This set of 999 values gives the reference distribution of the spatial scan statistic. The $p$ value of the cluster found in the original realization is given by the rank of the spatial scan statistic relative to the values in the reference distribution.

Figure 2 (left column) shows histograms of draws from the posterior distribution of $\rho$ and $\sigma$ obtained from fitting Model II to a single simulated realization with $\rho = 50$, $\sigma = 0.14$. Notice that there is quite a lot of uncertainty in the estimation of $\rho$.

Figure 3 shows the results of the modified procedure. We find that the proportion of $p$ values less than, for example, 0.05, is substantially smaller than the proportion obtained with the original spatial scan statistic, showing that the modified procedure is effective in reducing the number of false positives. The proportion is still larger than the nominal level. We believe this is due to the high uncertainty in our estimates of $\rho$, as the number of false positives found by the original spatial scan statistic depends strongly on $\rho$ (Figure 1).

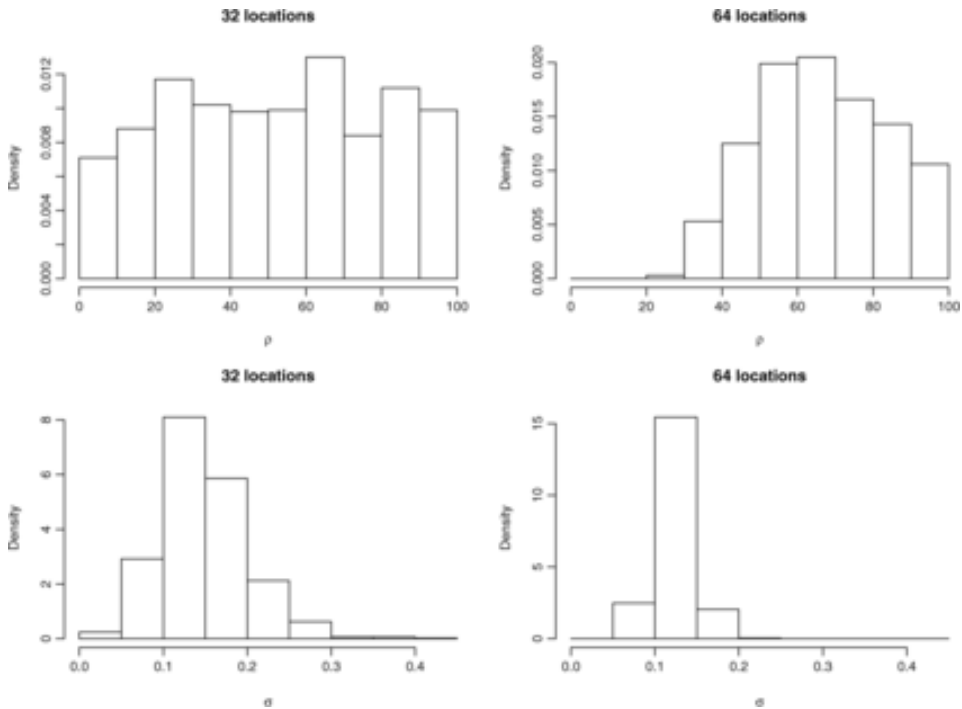

Fig. 2. *Histograms showing the posterior distributions of $\rho$ and $\sigma$ obtained from fitting Model* II *to one realization of the data with $\rho = 50, \sigma = 0.14$ (left column). The histograms on the right are similar posterior distributions using a dataset with 32 additional randomly placed locations and an incidence rate that is five times higher.*

We constructed reference distributions using the true parameter values and computed the modified $p$ values using these reference distributions. The results found here can be used as a guide for the performance of the modified spatial scan statistic algorithm in situations where we can obtain good parameter estimates. Figure 4 suggests that, with good parameter estimates, the proportion of small $p$ values does become very close to the expected level. The amount of variability in the curves in the figure provides an indication of the amount of uncertainty in the distribution of the spatial scan statistic, which is estimated by simulation.

The uncertainty in the estimation of $\rho$ would be reduced if there were more locations in the dataset and/or if the incidence rate were higher. As an example, we added an additional 32 locations randomly and generated a realization using $\rho = 50$ and $\sigma = 0.14$ with an incidence rate that is five times higher. The posterior distributions of $\rho$ and $\sigma$ obtained from fitting Model II are shown in Figure 2 (right column). Notice that the variability is much reduced. For example, the standard deviation for $\rho$ drops by 40%. For a given dataset, the high uncertainty in estimating $\rho$ may also be reduced by using more robust MCMC methods for fitting the spatial generalized linear mixed model [Christensen, Roberts and Sköld (2006)]. We did not explore this possibility in this paper. Our study shows that even in situations when conditions for estimating the parameters are not ideal, the modified spatial scan statistic procedure is nevertheless able to reduce the number of false alarms.

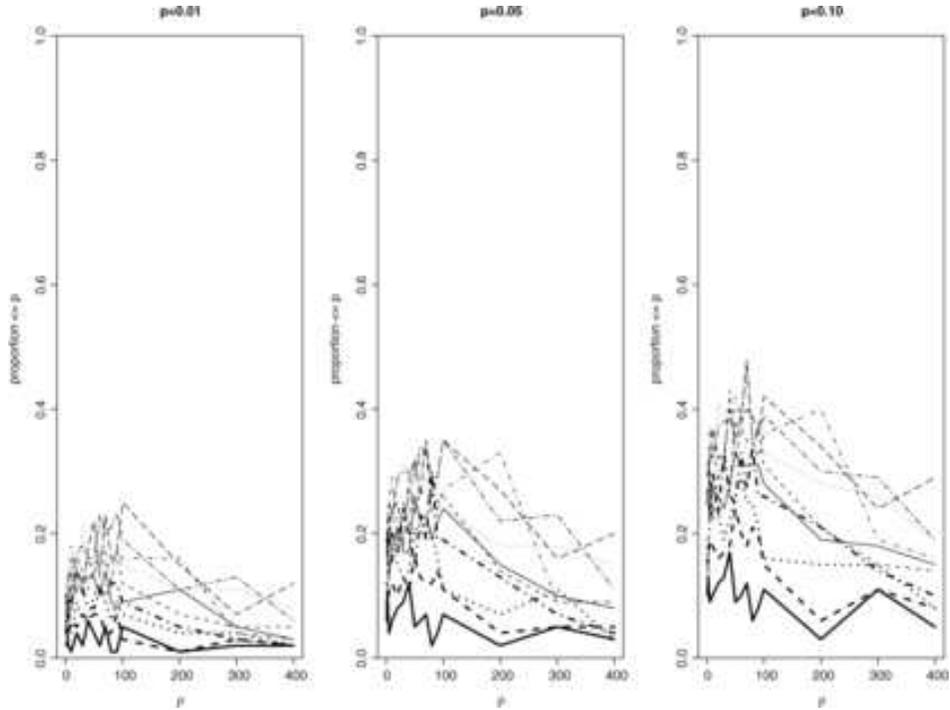

FIG. 3. *Plots of the proportion of p values equal to 0.01, 0.05 or 0.1 (left to right) or less, using the modified procedure, with the p values obtained by comparing the spatial scan statistic to a reference distribution generated with parameters estimated from a fitted generalized linear mixed model. The different line types refer to different values of $\sigma$ like in Figure 1.*

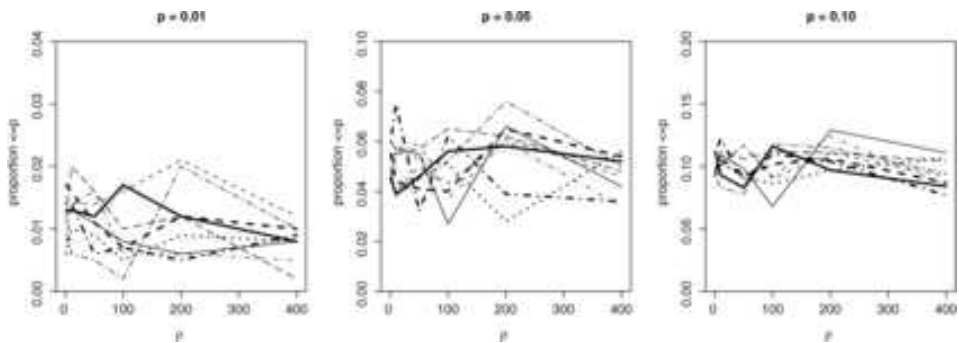

FIG. 4. *Plots of the proportion of p values equal to 0.01, 0.05 or 0.1 (left to right) or less, using the modified procedure, with the p values obtained by comparing the spatial scan statistic to a reference distribution generated with the true parameter values. The line types represent different values of $\sigma$ like in Figure 1.*

The priors we use for the parameters in this section and in Section 4 are the default in the geoRglm package: flat priors for $\beta$ and $\sigma$ and a uniform discrete prior on $(1, U)$ for $\rho$. When there is high uncertainty in the estimation of $\rho$, the choice of the upper bound $U$ may be important. In practice, for a large enough observation region, an upper bound roughly equal to the length of the region would usually be sufficient. The posterior draws of $\rho$ should also be examined, with a high proportion of values near the upper bound providing an indication that the upper bound is not large enough.

## 4. Data examples.

4.1. *New Mexico brain cancer data.* Here, we consider the analysis of a simple dataset to illustrate the performance of the modified spatial scan statistic. Kulldorff et al. (1998a) applied the scan statistic to a New Mexico brain cancer dataset that contains population and brain cancer incidence data for counties in New Mexico from 1973 to 1991. There were a total of 1175 cases. For each case, information is known about the county of residence, year of diagnosis, age, race and sex. The raw data can be obtained from the National Cancer Institute or from the SatScan software website http://www.satscan.org. The coordinates recorded for the 32 counties range from 8 to 162 for both the East–West and North–South directions. This dataset is of interest partly because there was concern in the Los Alamos community in 1991 about an apparent increase in brain tumor deaths. Kulldorff et al. (1998a) found a nonsignificant cluster with a $p$ value of 0.07, adjusting for age, sex and race.

Here, we consider the brain cancer data as an illustrative example of how our method can adjust for spatial correlation. Covariates can capture some of the spatial correlation in the data, partially taking the place of **Z** in (3). If covariates are not included, we found that the spatial scan statistic detects a highly significant cluster with a $p$ value of 0.009. This cluster consists of seven counties: Bernallillo, San Miguel, Sandoval, Sante Fe, Socorro, Torrance and Valencia.

To show the performance of the modified spatial scan statistic, we choose to ignore the covariates in the procedure, so that there is spatial correlation in the data that is unaccounted for. We first fit Model II to the yearly data and examine the trend, if any, of the various parameter estimates. Plots of the estimates (mean of the posterior distribution) and their 50% posterior credible regions are given in Figure 5. Most of the estimates are very similar from year to year, with overlapping 50% posterior credible regions. Note that the estimator of $\rho$ for the year 1987 is higher than the rest, though the difference is not significant. This year is within the nonsignificant but most likely space–time cluster of 1985 to 1989 found from the analysis of Kulldorff et al. (1998a), and the excess incidence is in Los Alamos from 1986 through 1989.

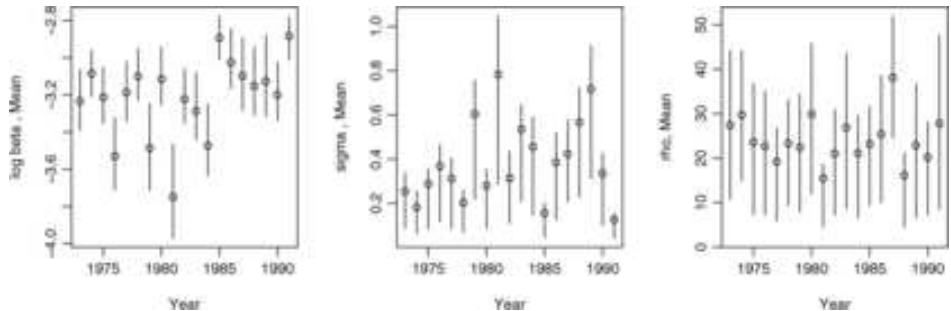

Fig. 5. *Plots of parameter estimates and their 50% posterior credible region of the generalized linear mixed model fit to the yearly New Mexico brain cancer data.*

Based on the results of Kulldorff et al. (1998a) and our analysis on the yearly data, we split the data into two sets, one from 1973 to 1982, and the other from 1983 to 1991, and consider the total occurrences in the counties during these two time periods. Thus, we roughly treat 10 years as one time unit. We use data from 1973–1982 as normal data for estimating the parameters of Model II, and data from 1983–1991 as test data. We fit Model II to the 1973–1982 data, with the Matérn model (5) for $C_\theta$, flat priors for $\beta$ and $\sigma$, and a uniform discrete prior on $(1, 70)$ for $\rho$. We take the posterior means $\beta = -0.834, \sigma = 0.176$, and $\rho = 20.94$ as our estimates.

We used the above parameter estimates in the Monte Carlo procedure to account for any spatial correlation in the data. (Recall that the original procedure computed the $p$ value by assuming that $\mathbf{Z} \equiv 0$ in Model II.) Specifically, using the fitted model, we simulate 999 new incidence datasets and compute the scan statistic for each of them. These spatial scan statistic values are then compared with the value obtained from the real data to obtain a $p$ value. Using this method, we find that the $p$ value for 1983–1991 data becomes 0.25. Thus, by using the modified spatial scan statistic, we find that the cluster is actually not significant.

In this illustrative example, we showed how modifying the spatial scan statistic by using Model II instead of Model I can help capture the spatial correlation due to unmeasured covariates. Of course, in a real application, any measured covariates should be included in the analysis. However, it is common for important covariates to be unmeasured and the modified scan statistic is a general procedure that can account for that.

4.2. *France chickenpox incidence data.* The incidence rates of 15 different diseases (e.g., influenza symptoms, chickenpox and diarrhoea) are recorded on a weekly basis for the 22 administrative regions of France. These regions span roughly 800 km for both the East–West and North–South directions. For some of the diseases, the data date back to as far as 1985 and data for most of these diseases are still being collected. The data collection is part of a comprehensive system set up to monitor the incidence of these diseases and to facilitate the detection of outbreaks. The data is collected from submissions by doctors participating in the network, and can be accessed at http://www.b3e.jussieu.fr/sentiweb. More details about this surveillance network can be found in Boussard et al. (1996). Deguen, Chau and Flahault (1998) indicated that some other covariates such as age and gender are also recorded. However, this additional data do not appear to be available from the above website.

Sampling bias could be an issue—since submission of data is voluntary, there might be undercoverage in areas where fewer doctors are in the network. In any region, the number of doctors and the actual doctors participating may change over time. No information about this appears to be recorded, other than a map showing participating doctors. However, the system computes an incidence rate from the data it receives. We assume that this calculation already takes into account the undercoverage problem and will ignore these issues in our analysis.

In this section we will perform our analysis on both the yearly and 4-weekly totals of the incidence of chickenpox (variella) for the period between 1996 to 2005. Chickenpox is contagious and we expect some degree of spatial correlation between the incidence rates of neighboring regions. The position of the administrative center of each region is used as the position for all cases occurring in that region. Deguen, Chau and Flahault (1998) looked at the epidemiology of chickenpox in France, finding summary statistics for the national time series of chickenpox incidence data. Deguen, Thomas and Chau (2004) built a SEIR model to estimate the contact rate of chickenpox. For influenza cases, Costagliola et al. (1991) and Viboud et al. (2004) built regression models based on nonepidemic data with the aim of using these models to identify future incidences of epidemics. In each of these cases, the spatial information of the data is not considered. We focus our analysis on the use of scan statistics to identify hotspots in both space and time. In particular, for chickenpox, we find that higher incidence rates may occur at different times for different locations. To our knowledge, this is the first attempt to identify both spatial and temporal outbreaks of chickenpox. In a retrospective study, these identified spatial–temporal outbreaks combined with other covariates can be used to study the socio-demographic factors that may contribute to the outbreaks and evaluate the effectiveness of certain intervention procedures such as mass immunization.

We first compute the regular scan statistic for the data over this period. In particular, the SatScan program can perform a prospective analysis using data from prior time periods to detect clusters in the current period. Thus, we use data in time period 1 as a baseline to detect clusters in time period 2, data in time periods 1–2 to detect clusters in time period 3, etc.

For yearly total incidence data, we find that with the regular spatial scan statistic, clusters with $p$ values of 0.001 were found for all the years from 1997 to 2005, an indication that the independent Poisson model is not appropriate. For the 4-weekly data from 1997 to 2005, the regular spatial scan statistic identified 113 primary clusters out of 117 4-week periods, 109 of which have $p$ values 0.05 or less and 108 have $p$ values 0.001 or less. For this data, the regular spatial scan statistic identified such an excessive number of outbreaks that it is not practically useful.

We next fit the spatial generalized linear mixed Model II to the chickenpox incidence rate for an initial period and then use the fitted model to do a Monte Carlo adjustment to the scan statistic as described in Section 3 for subsequent time periods. With the 1996 data, we obtained posterior mean estimates of 6.6, 572 and 0.4 for $\beta, \rho$ and $\sigma$ for Model II using the Matérn model for the covariance function with $\nu = 0.5$, and with flat priors for $\beta$ and $\sigma$ and a uniform discrete prior on (1,700) for $\rho$. Plots of the posterior draws of $\beta, \rho$ and $\sigma$ are shown in Figure 6. Note that in the posterior sample of $\sigma$, more than 95% of the draws are greater than 0.10, suggesting that Model I, with $Z \equiv 0$, is not appropriate. Formal testing may be done, for example, with Bayes factors [Kass and Raftery (1995)].

We then computed the spatial scan statistic for each year from 1997 to 2005 and compared each of these statistics with the spatial scan statistics computed from simulated realizations using the estimates of $\beta, \rho$ and $\sigma$. With this procedure, the modified $p$ values look more reasonable, and in particular, only 1999 and 2004 have $p$ values less than 0.05. Figure 7 shows maps of the incidence counts per 10 million. The primary (solid line) and first secondary (dashed line) clusters identified by the spatial scan statistic are also indicated. Note that although relatively high values occur somewhere in each year, the most significant ones are clearly those corresponding to 1999 and 2004.

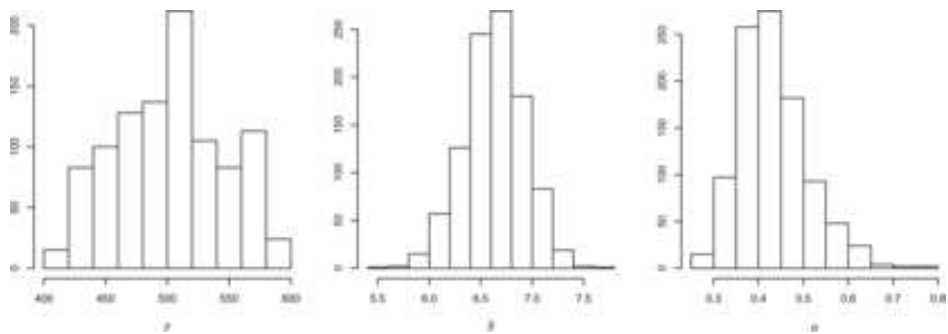

FIG. 6.   *Plots of the posterior draws of $\beta$, $\rho$ and $\sigma$, obtained from fitting Model* II *to the 1996 chickenpox data.*

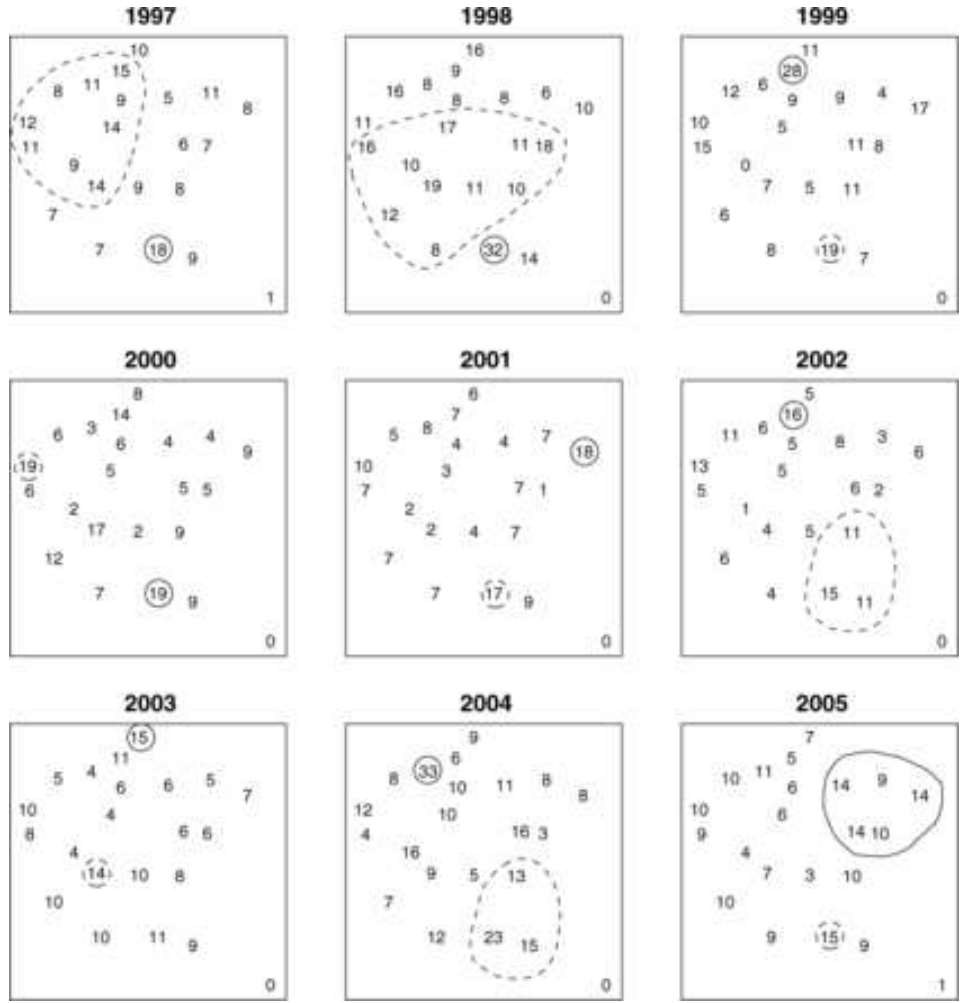

FIG. 7. *Maps of incidence rates of chickenpox for 1997 to 2005 in the 22 administrative regions of France. The primary and first secondary clusters found by the spatial scan statistic are indicated (solid and dashed lines, resp.).*

We repeat the same procedure for the 4-week data, fitting a spatial GLM to each 4-week period of 1996, using the same priors as above for $\beta, \sigma$ and $\rho$. We then used the overall mean of the estimates to simulate realizations according to Model II. These were found to be 3.8, 390 and 1.05, respectively for $\beta, \rho$ and $\sigma$. The estimate of $\sigma$ is larger here, reflecting the increased variability in this data compared to the yearly data. From Figure 1, we expect the modified $p$ values to be much larger. Indeed, we find that this is the case. Out of 117 4-weekly periods, only 6 have $p$ values 0.1 or less. Of these, only 2 have $p$ values 0.05 or less.

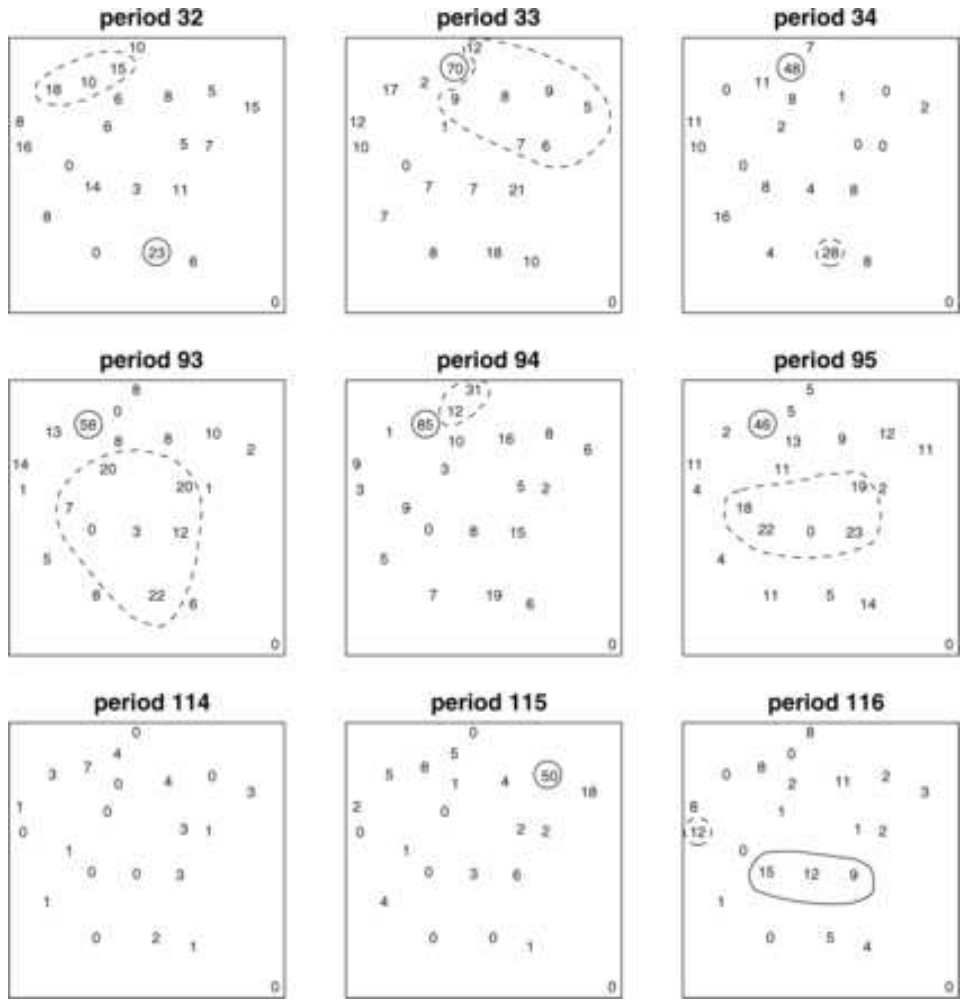

FIG. 8. *Maps of incidence rates of chickenpox in the 22 administrative regions of France for the 3 most significant 4-week periods found in the 1997–2005 data, periods 33, 94 and 115. Also shown are maps for the periods just before and after these significant periods. The primary and first secondary clusters found by the spatial scan statistic are indicated (solid and dashed lines, resp.).*

Figure 8 shows incidence maps (counts per 1 million) for the three most significant 4-week periods, periods 33, 94 and 115 with $p$ values 0.06, 0.04 and 0.04, respectively. Also shown are the 4-week periods just before and after the significant periods. The two main clusters are identified as before. Comparison among the three consecutive periods shows that the significant clusters found by our modified procedure appear to be hotspots in both space and time.

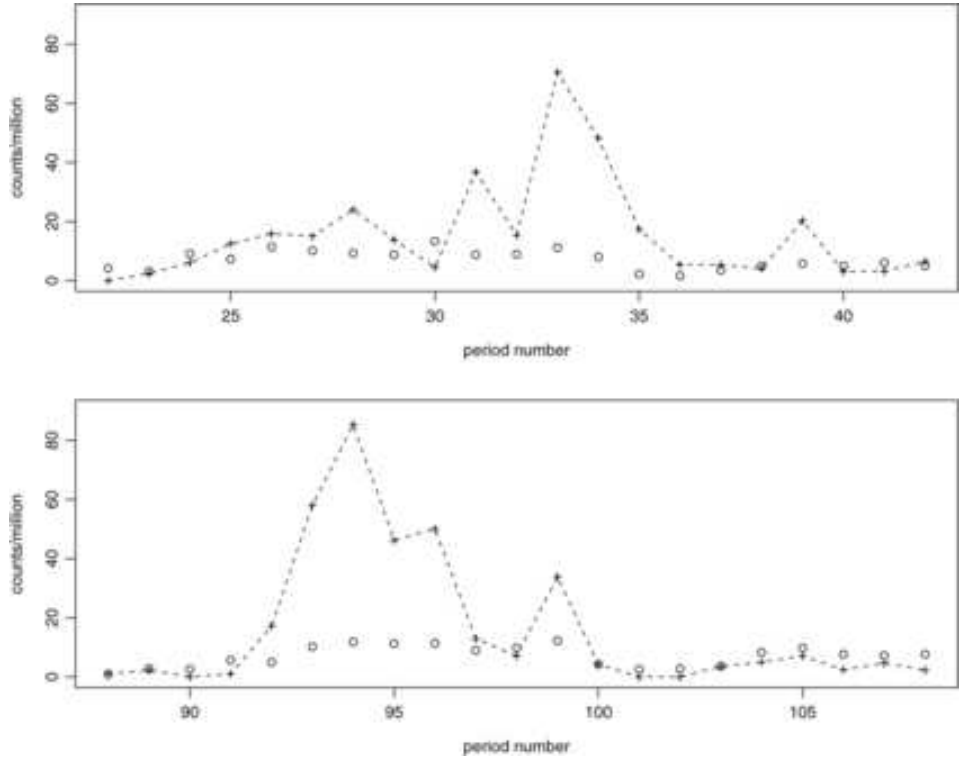

Fig. 9. *Time series of chickenpox cases (counts per million; plus signs) at the two locations that correspond to the significant clusters found at periods 33 and 94 by the modified spatial scan statistic procedure. Also shown are the average counts over all the regions.*

The maps in Figure 8 suggest occurrence of an outbreak that rises to a peak and then dies off. Figure 9 shows time series of chickenpox cases (counts per million, indicated by the plus signs) at the two locations that yielded the significant clusters at periods 33 and 94. The circles in the plots show the average counts over all the regions. Our method appears to identify when the peak occurs, but we did not extensively test this.

The above analysis does not take into account multiple comparisons in time. Thus, there may be significant clusters detected even if there are no real clusters due to testing in multiple time periods. A way to deal with this is to apply FDR techniques [Benjamini and Hochberg (1995)]. Here, we use

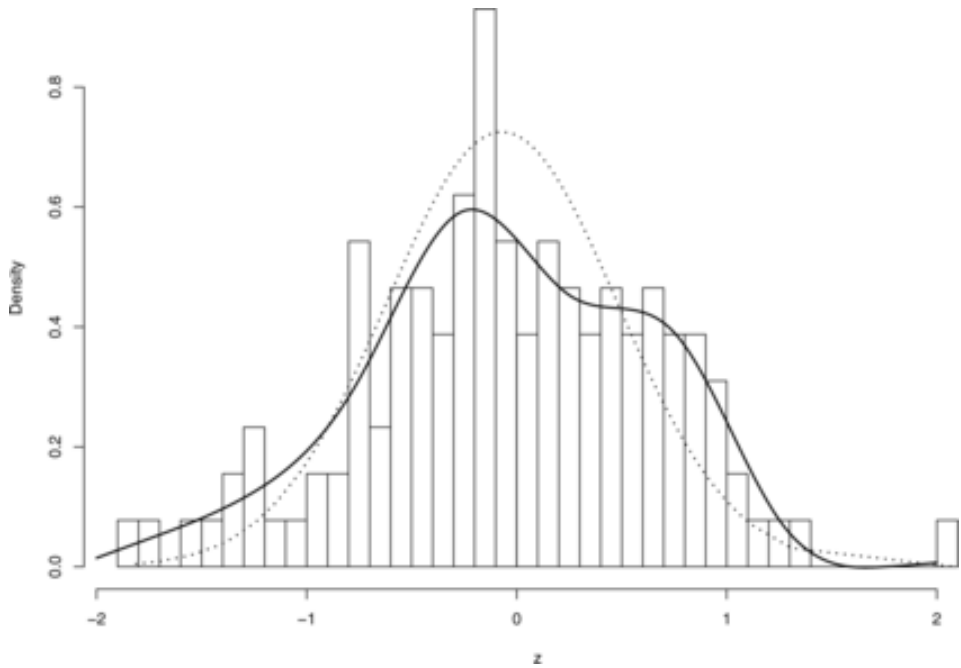

Fig. 10. *Histogram of the transformed p values, together with the fitted density (solid line) and density of the empirical null (dashed line).*

the local FDR version employed in Efron (2004), using the empirical null to compute the FDR.

Specifically, we take the 129 $p$ values obtained after applying the modified spatial scan statistic and transform them to $z$ values, that is,

$$z_i = \Phi^{-1}(p_i), \qquad i = 1, \ldots, 129,$$

where $\Phi$ is the standard normal cumulative distribution function. A histogram of these $z$ values is shown in Figure 10. Negative values of $z$ indicate possible clusters.

The empirical density $f(\cdot)$ is obtained by fitting a natural cubic spline to the $z$ values. Assuming that the empirical null $f_0(\cdot)$ is normal, we find that its mean and standard deviation are, respectively, $\delta_0 = -0.07$ and $\sigma_0 = 0.55$. The reader is asked to refer to Efron (2004) for details on the fitting procedure. Figure 10 also shows plots of $f$ and $f_0$ (solid and dashed lines, resp.). The value of the FDR is then taken to be

$$fdr(z) = f_0(z)/f(z).$$

This is an upper bound of the probability $P(\text{Uninteresting}|z) = p_0 f_0(z)/f(z)$, where $p_0$ is the proportion of uninteresting cases. Here, we assume that $p_0$ is large, say, $p_0 > 0.9$.

The histogram of Figure 10 suggests the possibility that the null is a mixture of two distributions. This may be due to the seasonality present in the data. However, there appears to be spatial variation in seasonality, with peaks occurring at different times at different locations, making it difficult to estimate the two components of the mixture distribution. Assuming that the null distribution is normal, we find that the FDR for periods 33, 94 and 115 are respectively 0.26, 0.12 and 0.17. Using 0.1 as the cut-off for the FDR, none of the three detected clusters are significant, although period 94 is close to the cut-off. We note that if the null distribution is indeed a mixture of two distributions, the true FDR would be smaller, likely making some of the detected clusters significant.

As a validation analysis, we fit time series models to the data at each individual location, and then apply outlier detection methods [Tsay (1986)] to identify hotspots in time. Specifically, we fit ARMA models to the logarithm of the data. A value of 0.5 was added to zeros in the data before taking logs. For most of the locations, seasonality was detected at lag 13, corresponding to yearly seasonal effects. A total of 12 hotspots at individual locations were found, at time periods 12, 23, 41, 51, 75, 77, 89, 92, 98, 99, 113 and 115.

Of these 12 hotspots, the last one, at time period 115, corresponds to the hotspot identified by the modified scan statistic. Since the outlier detection method does not include any spatial multiple comparison adjustment, it finds more hotspots. Even so, it did not find the hotspots at time periods 33 and 94 in Figure 8 because the method does not incorporate any spatial information.

We note that three of the 12 hotspots agreed exactly with those found by the regular spatial scan statistic. In four other cases, the regular spatial scan statistic found clusters containing multiple locations, which the time series method obviously cannot detect. Of these, two of the clusters found by the regular spatial scan statistic contained the location identified by the outlier detection method. Of the remaining 5 temporal hotspots, there were higher rates at other locations that the spatial scan statistic identified instead. Thus, including spatial information provides additional flexibility in the detection of hotspots. However, the regular spatial scan statistic, which does not take into account spatial correlation, finds a hotspot at almost every time period, too many to be useful. The modified scan statistic, by accounting for spatial correlation, substantially reduces the number of identified hotspots. In addition, by applying FDR techniques, we can also account for multiple comparisons in time. The modified spatial scan statistic and the time series outlier detection method can be used as complementary methods to study the spatial and temporal aspects of a dataset.

Note that in our analysis we used actual physical distances between administrative centers. Two physically distant regions might be closely correlated if they have extensive transportation services. We did not consider this in our analysis since it is not clear how distances should be re-defined in this case.

Finally, the definition of hotspots is dependent on the model assumption of the process without hotspots. Information about the region sizes, population densities and population distribution in each region, as well as other covariates, might be useful in building a more structured model. More hotspots may be identified if a more structured model that incorporates these other factors is fit to the data. This information, however, was not available to us.

**5. Conclusion.** The spatial scan statistic introduced by Kulldorff (1997) can be used to detect hotspots in the incidence of disease and has been widely used in epidemiological studies. The null model of the spatial scan statistic is Poisson counts at each location, with spatially independent rates. The Poisson assumption makes strong statements about the variability of the data which may not hold in practical applications. Overdispersion can occur through unrecorded covariates that are related to the disease or through errors in covariate measurement. Furthermore, there is often reason to believe that there is spatial correlation between subregions beyond that measured by covariates.

We showed in our simulation study that the spatial scan statistic can report a high proportion of small $p$ values when overdispersion or spatial correlation exists in the data. This results in many false alarms in disease detection. We showed how one can easily account for this by means of a simple procedure, specifically, by obtaining the distribution of the spatial scan statistic from simulating a generalized spatial linear model (Model II) instead of the independent Poisson model (Model I). This new procedure is only slightly more computationally intensive than the original procedure, but can greatly reduce the number of false alarms. Kulldorff (1997) also considered the binomial model. In this paper we focus on the Poisson model, but we expect the findings to be similar for the binomial model.

There may be slight loss of power in this modified procedure—we will investigate this in a future work. However, from our data analyses in Section 4, we find that the most significant clusters are still detected with our procedure. Also, with fewer clusters identified, it will be easier to further investigate the identified clusters. With the classical spatial scan statistic, a lot of resources could be spent weeding out the false alarms.

## APPENDIX

PROOF OF PROPOSITION 2. Letting $U_i = \sqrt{n} Z_i \sim N(0, \sigma^2)$, it follows from the Taylor expansion that

$$e^{Z_i} = 1 + U_i n^{-1/2} + \tfrac{1}{2} U_i^2 n^{-1} + O_p(n^{-3/2}),$$

$$\lambda_A = e^\beta \sum BP_i e^{Z_i}$$
$$= e^\beta \sum BP_i(1 + U_i n^{-1/2} + \tfrac{1}{2}U_i^2 n^{-1}) + O_p(n^{-3/2})$$
$$= \lambda_A^* + e^\beta \sum BP_i(U_i n^{-1/2} + \tfrac{1}{2}(U_i^2 - 1)n^{-1}) + O_p(n^{-3/2})$$
$$= \lambda_A^* + \delta,$$

where $\lambda_A^* = e^\beta \sum BP_i(1 + \frac{1}{2n})$. It can be shown that $\lambda_A^* = \mathrm{E}[\lambda] + O(n^{-3/2})$, $\delta = O_p(n^{-1/2})$, $\mathrm{E}[\delta] = O(n^{-3/2})$ and $\mathrm{E}[\delta^2] = e^{2\beta} V_n n^{-1}$.

Let $f_j(x) = e^{-x}\frac{x^j}{j!}$. The Taylor expansion of $f_j(\lambda_A)$ at $\lambda_A^*$ gives

$$f_j(\lambda_A) = f_j(\lambda_A^*) + f_j'(\lambda_A^*)\delta + f_j''(\lambda_A^*)\delta^2/2 + O_p(n^{-3/2})$$
$$= f_j(\lambda_A^*)\left\{1 + \left(\frac{N}{\lambda} - 1\right)\delta + \left(\frac{N^2 - N}{\lambda^2} - \frac{2N}{\lambda} + 1\right)\frac{\delta^2}{2}\right\} + O_p(n^{-3/2})$$

and

$$\sum_{j=k}^\infty P_j^{(2)} = \sum_{j=k}^\infty \mathrm{E}[f_j(\lambda_A)]$$
$$= \sum_{j=k}^\infty \left\{f_j(\lambda_A^*)\left(1 + \left(\frac{N^2 - N}{\lambda^2} - \frac{2N}{\lambda} + 1\right)e^{2\beta}\frac{V_n}{2n}\right)\right\} + O_p(n^{-3/2})$$
$$= \sum_{j=k}^\infty P_j^{(1)} + \sum_{j=k}^\infty (P_{N-2}^{(1)} - 2P_{N-1}^{(1)} + P_j^{(1)})e^{2\beta}\frac{V_n}{2n} + O_p(n^{-3/2})$$
$$= \sum_{j=k}^\infty P_j^{(1)} + (P_{k-2}^{(1)} - P_{k-1}^{(1)})e^{2\beta}\frac{V_n}{2n} + O_p(n^{-3/2}). \qquad \square$$

Department of Statistics  
Columbia University  
New York, New York 10027  
USA  
E-mail: meng@stat.columbia.edu

Department of Statistics  
and Operations Research  
University of North Carolina at Chapel Hill  
Chapel Hill, North Carolina 27599  
USA  
E-mail: zhuz@email.unc.edu